\documentclass{article}
\usepackage[ansinew]{inputenc}
\usepackage{latexsym}
\newcommand{\be}{\begin{equation}}
\newcommand{\ee}{\end{equation}}

\newcommand{\dt}{\mbox{\boldmath$:$}}
\newcommand{\vdt}{\mbox{\bf$\vdots$}}

\newcommand{\no}{\noindent}

\begin{document}

\title{The Thirring interaction in the two-dimensional axial-current-pseudoscalar derivative coupling model}

\author{\small{L. V. Belvedere$^\ast$ and A. F. Rodrigues$^{\ast\ast}$}\\
\small{\it{Instituto de F\'{\i}sica - Universidade Federal Fluminense}}\\
\small{\it{Av. Litor\^anea S/N, Boa Viagem, Niter\'oi, CEP 24210-340}}\\
\small{\it{Rio de Janeiro, Brasil}}\\
\small{$\ast$\,\it{belve@if.uff.br}}\\
\small{$\ast\ast$\,\it{armflavio@if.uff.br}}
}
\date{\today}

\maketitle

\begin{abstract}
\small{We reexamine the two-dimensional model of massive fermions interacting with a massless pseudoscalar field via axial-current-pseudoscalar derivative coupling. Performing a canonical field transformation on the Bose field algebra the model is mapped into the Thirring model with an additional vector-current-scalar-derivative interaction (Schroer-Thirring model). The complete bosonized version of the model is presented. The bosonized composite operators of the quantum Hamiltonian are obtained as the leading operators in the Wilson short distance expansions. }
\end{abstract}



\section{Introduction}

\setcounter{equation}{0}

The two-dimensional spinor-scalar model with derivative couplings has been the subject of various investigations 
within different approaches \cite{Schr,RS,1,2,3,4,5,6,7}. The model describing a Fermi field interacting via derivative coupling with two Bose fields, one scalar and the other pseudoscalar, was analyzed in Refs. \cite{4,5,6}. For a certain choice of the coupling parameters the equivalence between the fermionic sector of the derivative coupling model and the Thirring model can be established in a weak form between the fermionic Green's functions of the corresponding models. However, this weak equivalence only works under the expense of introducing opposite metric quantization for the bosonic fields \cite{5,6}, or by considering one derivative interaction term with imaginary coupling 
parameter \cite{4}. As a matter of fact, this is the only way under which the degrees of freedom in the two models can be artificially matched. In Ref. \cite{5} the connection between the two models is analyzed within the operator formulation. In order to establish the isomorphism between the fermionic Green's functions of the two models, the Bose fields are considered with oposite metric and a special combination of the original {\it{three}} bosonic degrees of freedom is introduced to define {\it{two}} new bosonic fields. In this way the operator solution is given in terms of both a spurion field and the Thirring field operator. Moreover, the
transformation performed in Ref. \cite{5} is meaningless in the model with massive fermions since it presupposes that the bosonic fields are free and massless; the spurion 
field has no definite parity and spoils the mass operator.

The model describing a massless pseudoscalar field interacting via derivative coupling with fermions (massless Rothe-Stamatescu model) \cite{RS} was considered in Ref. \cite{7} within the smooth bosonization approach. The similarity between the massless Rothe-Stamatescu model and the Thirring model is suggested. Within this approach, the claimed similarity between the two models follows from the fact that the Lagrangian of the Thirring model modified by the introduction of an auxiliary vector field is ``almost'' the Lagrangian of the massless Rothe-Stamatescu model except for the existence of a scalar field with indefinite metric. However, this is a naïve conclusion and does not implies neither the equivalence between the two models, nor the presence of the Thirring interaction in the massless Rothe-Stamatescu model. The structural aspects of the bosonization of the Thirring model with the use of an auxiliary vector field are discussed in Ref. \cite{bel}. The use of an auxiliary vector field to reduce the action of the Thirring model to a quadratic action in the Fermi field introduces a redundant Bose field algebra which contains more degrees of freedom than those needed for the description of the model. It is shown in Ref. \cite{bel} that the only effect of the redundant decoupled Bose fields is to generate constant contributions to the Wightman functions in the Hilbert space of states. In contrast to what occurs in two-dimensional gauge theories \cite{LS,bel1}, the spurious Bose field combination has no physical consequences, since it carries no charge selection rule, and reduces to the identity in a positive metric Hilbert space of states \cite{bel}. 

From our point of view, the approach followed in Refs. \cite{5,6} to establish the weak equivalence between the Thirring model and the derivative coupling (DC) model,  does not exhibit the true physical properties of the complete Hilbert space of the model. The  relation between the Thirring model and the derivative coupling model has never been very clear because of an incomplete understanding of the actual role played by the fermionic quartic self-interaction in the derivative coupling model. A clear demonstration at the operator level of the role played by the Thirring interaction underneath the derivative coupling model is lacking in the literature. The purpose of the present work is to fulfill this gap. To this end we shall consider the two-dimensional model of a massless pseudoscalar field interacting via derivative coupling with a massive Fermi field. This model corresponds to the Rothe-Stamatescu model \cite{RS} in the zero mass limit for the boson field and modified to include a mass term for the Fermi 
field. The model describes a massless pseudoscalar field interacting with a massive Fermi field via axial-current-pseudoscalar derivative coupling. Throughout this paper we shall refer to this modified Rothe-Stamatescu model as MRS model. The main purpose of the present paper is to analyze the MRS model within the operator formulation in order to explicitly show that the MRS model is equivalent to the Thirring model with an
additional vector-current-scalar derivative interaction, i. e., the Schroer-Thirring model. The hidden Thirring interaction in the MRS model is exhibited compactly by performing a canonical transformation on the Bose fields. The operator solution for the quantum equations of motion of the MRS model is written in terms of the Mandelstam Fermi field operator of the Thirring model interacting with a scalar field via derivative coupling. The charge sectors of the MRS model are mapped into the charge sectors of the massive Thirring model. The paper is organized as follows: In section $2$ we present the operator formulation to display the Thirring interaction in the MRS model. The equivalence of the MRS model with the Schroer-Thirring model is established at the operator level. In section $3$ the complete bosonized version of the model is presented. The bosonized composite operators of the quantum Hamiltonian are computed as the leading operators in the Wilson short distance expansion for the operator products at the same point. The conclusion and final comments are presented in section $4$.

\section{The mapping of the MRS model into the Schroer-Thirring model}

The two-dimensional model describing a massive Fermi field interacting with a massless pseudoscalar Bose field via axial-current derivative coupling is defined by the classical Lagrangian density \footnote{The conventions used are: 
$$ \gamma^0 = \pmatrix{0 & 1 \cr 1 & 0}\,\,,\,\,\gamma^1 = \pmatrix{0 & 1 \cr -1 & 0}\,\,,\,\,\gamma^5 = \gamma^0 \gamma^1\,\,,\,\,\psi = \pmatrix{\psi_1 \cr \psi_2}\,,$$

$$\,g^{00} = - g^{11} = 1\,,\,\epsilon_{\mu\nu} = - \epsilon_{\nu\mu}\,,\,\epsilon_{01} = \epsilon^{10} = 1\,,\,\gamma^\mu \gamma^5\,=\,\epsilon^{\mu \nu}\,\gamma_\nu\,.
$$

\no For free massless Bose fields, $\phi (x) = \phi_{r} (x) + \phi_{\ell} (x)\,,\tilde\phi (x) = \phi_{r} (x) - \phi_{\ell} (x)\,\,\,,\,\,\,\phi_{r,\ell} = \phi (x^\mp )\,.$
}, 

\be\label{L1}
{\cal L} (x) = \bar \psi (x)\,\big ( i \gamma^\mu\,\partial_\mu\,-\,m_o\,\big )\,\psi (x)\,
+\,\frac{1}{2}\,\partial_\mu\,\tilde \phi (x)\,\partial^\mu\,\tilde \phi (x)\,
+\,g\,\Big (\,\bar \psi (x)\,\gamma^\mu\,\gamma^5\,\psi (x)\,\Big )\,\partial_\mu\,\tilde \phi (x)\,.
\ee

\no The Lagrangian (\ref{L1}) describes the Rothe-Stamatescu model \cite{RS} in the zero mass limit of the pseudoscalar field and modified to include a mass term for the Fermi field (MRS model). The quantum theory is defined by the following equations of motion 
   
\be\label{emf}
\big ( i \gamma^\mu\,\partial_\mu\,-\,m_o\,\big )\,\psi (x)\,=\,
\,g\, \gamma^\mu\gamma^5\,N [ \psi (x)\, \partial_\mu\,\tilde \phi (x)\,]\,,
\ee

\be\label{emp}
\Box \tilde \phi (x)\,=\,-\,g\,\partial_\mu\,
\vdt \Big (\,\bar \psi (x)\,\gamma^\mu\,\gamma^5\,\psi (x)\Big )\vdt\,.
\ee

\no The dots in Eq. (\ref{emp}) mean that the current is computed as the leading operator in the Wilson short distance expansion and the normal product in (\ref{emf}) is defined by the symmetric limit \cite{RS,AAR}

\be
N [ \psi (x) \partial_\mu \tilde\phi (x) ] \doteq \lim_{\varepsilon \rightarrow 0}\,\frac{1}{2}\,\Big \{
\partial_\mu  \tilde\phi (x + \varepsilon ) \psi (x) + \partial_\mu \tilde\phi (x - \varepsilon ) \psi (x) \Big \}\,.
\ee

\no As a consequence of the axial-current-pseudoscalar derivative interaction, for massive fermions ($m_o \neq 0$) the field $\tilde\phi$ does not
remains free due to the non-conservation of the axial current \footnote{In the Schroer model \cite{Schr}, which describes a massive Fermi field interacting with a scalar field via vector-current-scalar derivative 
coupling, the scalar Bose field remains free due to the conservation of the vector current $\bar\psi\,\gamma^\mu\,\psi$.} in Eq. (\ref{emp}).

For massless Fermi fields the quantum model described by the Lagrangian (\ref{L1}) (massless Rothe-Stamatescu model) is a scale invariant theory with anomalous scale dimension \cite{RS}. As in the standard Thirring model \cite{Swieca}, in order that the theory described by the Lagrangian (\ref{L1}) has the model with a massless fermion as the short distance
fixed point the scale dimension of the mass operator must be

\be\label{dpsi1}
D_{_{\bar \psi \psi}} < 2\,.
\ee

\no In what follows the mass term should be understood as a perturbation in the scale invariant model. 

The operator solution for the quantum equations of motion is given in terms of Wick-ordered exponentials \cite{RS,5,AAR},

\be\label{psi}
\psi (x)\,=\,{\cal Z}_\psi^{- \frac{1}{2}} \,\dt e^{\,\textstyle i\,g\,\gamma^5\,\tilde \phi (x)}\dt\,\psi^{(0)} (x)\,,
\ee

\no where ${\cal Z}_\psi$ is a wave function renormalization constant \cite{RS,AAR} and $\psi^{(0)}$ is the free massive Fermi field,

\be
\big (i\,\gamma^\mu \partial_\mu - m_o \big )\, \psi^{(0)} (x) = 0\,.
\ee

\no The bosonized expression for the free Fermi field operator $\psi^{(0)}$ is given by the Mandelstam field operator \cite{M},

\be
\psi^{(0)} (x) = \Big (\frac{\mu}{2 \pi} \Big )^{1/2}\,
e^{\,\textstyle -\,i\,\frac{\pi}{4}\,\gamma^5}\,
\dt e^{\textstyle\,i\,\sqrt \pi\,\{\,\gamma^5\,\tilde \varphi (x)\,+\,\int_{x^1}^{\infty}\,
\partial_0\,{\tilde\varphi} (x^0,z^1)\,d z^1\,\}}\dt\,,
\ee

\no where $\mu$ is an infrared regulator reminiscent of the free massless theory. For $m_o = 0$, the field $\tilde\varphi$ is free and massless such that,

\be
\epsilon_{\mu \nu} \partial^\nu \tilde \varphi (x) = \partial_\mu \varphi (x)\,.
\ee

\no The meaning of the notation $\dt ( \bullet ) \dt$ in the field operators is that the Wick ordering is performed by 
a point-splitting limit in which the singularities subtracted are those of the free theory. In this way, the  Wilson short-distance expansions are performed using the two-point function of the free massless field 

\be
[ \Phi^{(+)} (x)\, ,\, \Phi^{(-)} (0)]_{x \approx 0}\, =\,-\,\frac{1}{4 \pi}\,\ln \{- \mu^2 (x^2 \,+\, i\, \epsilon\, x^0)\}\,.
\ee

In order to preserve the classical symmetry of the model the vector 
current is computed with the regularized point-splitting limit procedure \cite{RS,AAR},

\be\label{cur}
{\cal J}^\mu (x) = \vdt \bar\psi (x) \gamma^\mu \psi (x) \vdt = \lim_{\varepsilon \rightarrow 0}\,
\,\Big \{\,\bar \psi ( x + \varepsilon ) \gamma^\mu\,
e^{\textstyle\,-\,i\,g\,\int_x^{x + \varepsilon }\,
\epsilon_{\mu \nu}\,\partial\,^\nu\,\tilde \phi (z)\,d z^\mu}\,\psi (x)\,
-\,V. E. V.\,\Big \}\,,
\ee

\no with the wave function renormalization constant ${\cal Z}_\psi$ given by \cite{RS,AAR},

\be
{\cal Z}_\psi (\epsilon)\,=\,e^{\,  g^2 [ \tilde \phi^{(+)} (x + \epsilon )\, ,\, \tilde\phi^{(-)} (x) ]}
\,.
\ee

\no The vector current is given by \cite{RS,AAR}

\be\label{vc}
{\cal J}^\mu (x) = j^\mu_{f} (x)\,-\,\frac{g}{\pi}\,\epsilon^{\mu\nu}\partial_\nu\,\tilde\phi (x)\,,
\ee

\no where $j^\mu_{f} (x)$ is the free fermion current,

\be
j^\mu_f (x) = \vdt \bar\psi^{(0)} (x) \gamma^\mu \psi^{(0)} (x) \vdt = - \frac{1}{\sqrt \pi}\,\epsilon^{\mu \nu}\,\partial_\nu \tilde \varphi (x)\,,
\ee

\no and the axial current is,

\be\label{5c}
{\cal J}_\mu^5 (x) = \epsilon_{\mu\nu} {\cal J}^\nu (x) = \,-\,\partial_\mu\,\Big (\,\frac{1}{\sqrt\pi}\,\tilde\varphi (x)\,+\,\frac{g}{\pi}\,\tilde\phi (x)\,\Big )\,.
\ee

\no With the expression (\ref{5c}) for the axial current one can write the quantum equation of motion (\ref{emp}) in the bosonized form

\be\label{bem}
\Big ( 1 - \frac{{g}^2}{\pi} \Big )\,\Box \tilde \phi (x)\,=\,\frac{g}{\sqrt \pi}\,
\Box \tilde \varphi (x)\,.
\ee

\no In order to avoid opposite metric quantization for the fields $\tilde\phi$ and $\tilde\varphi$, we shall consider the model defined for $g^2$ in the range \footnote{The massless RS model at the critical point $g^2 = \pi$ was considered in Ref. \cite{7} within the Hamiltonian formalism in the context of an enlarged gauge invariant theory.}

\be
\frac{g^2}{\pi} < 1\,.
\ee

\no We shall ignore the infrared problems of the two-dimendional massless free boson field since the selection rules carried by the Wick-ordered exponentials ensure the construction of a positive metric Hilbert space \cite{Wightman,Swieca}. The bosonized mass operator takes the form 

\be\label{bmo}
\vdt \bar \psi (x)\,\psi (x) \vdt\,=\,-\,\frac{\mu}{\pi}\,\dt \cos \,\Big (\,2 \sqrt \pi\,
\tilde \varphi (x)\,+\,2\, g\,\tilde \phi (x)\,\Big ) \dt\,.
\ee

\no From the bosonized mass operator (\ref{bmo}) and from the equation of motion (\ref{bem}) we see that for $m_o \neq 0$ the fields $\tilde\varphi$ and $\tilde\phi$ are sine-Gordon-like fields. In this case the fields $\tilde\varphi$ and  $\tilde \phi$ are free massless fields and the axial current (\ref{5c}) in the equation of motion (\ref{emp}) is conserved. In the original Rothe-Stamatescu model ($m_o = 0$) the pseudoscalar field $\tilde\phi$ is  massive and the axial current has an anomaly. In this case the field $\tilde\phi$ remains free in the presence of the axial-vector-pseudoscalar derivative interaction, aside from a finite mass and wavefunction renormalization \cite{RS}.

In order to have canonical commutation relation for the field $\tilde \phi$ we perform the field scaling

\be\label{fs}
\tilde \phi (x) = \Big ( 1 - \frac{{g}^2}{\pi} \Big )^{- \frac{1}{2}}\,\tilde \phi^\prime (x)\,.
\ee

\no After the field scaling (\ref{fs}), the mass operator (\ref{bmo}), the vector current (\ref{vc}) and the equation of motion (\ref{bem}) can be rewritten as

\be\label{e1}
\vdt \bar \psi (x)\, \psi (x) \vdt = \,
-\,\frac{\mu}{\pi}\,\dt \cos \,\Big (\,2 \sqrt \pi\,
\tilde \varphi (x)\,+\,\frac{2 g}{\sqrt{1 - \frac{{g}^2}{\pi}}}\,\tilde \phi^\prime (x)\,\Big ) \dt\,,
\ee

\be\label{e2}
{\cal J}_\mu (x)\,=\,-\,\frac{1}{\pi}\,\epsilon_{\mu \nu}\partial^\nu\,\Big ({\sqrt\pi} \tilde\varphi (x) + \frac{g}{\sqrt{1 - \frac{{g}^2}{\pi}}}\,\tilde\phi^\prime (x) \Big )\,,
\ee

\be\label{e3}
\Box \Big ( \sqrt\pi \tilde\phi^\prime (x) - \frac{g}{\sqrt{1 - \frac{g^2}{\pi}}} \tilde \varphi (x) \Big ) = 0\,.
\ee

\no The scale dimension of the mass operator is given by

\be\label{dpsi2}
D_{_{\bar \psi \psi}} = \frac{\beta^2}{4 \pi}\,,
\ee

\no with

\be
\beta^2 \doteq \frac{4 \pi}{1 - \frac{{g}^2}{\pi}}\,.
\ee

\no On account of (\ref{dpsi1}) and (\ref{dpsi2}), for short distances the mass perturbation becomes increasingly negligible for $g^2 < \pi / 2$. Notice that the scale dimension of the mass operator is the same as that of the massive Thirring model with coupling parameter $g$. In terms of the field $\tilde \phi^\prime$ the Fermi field operator can be rewritten as

\be
\psi (x) = {\cal Z}_\psi^{- 1 / 2}\,\dt \exp\,\Big \{\, i\,\frac{g}{\sqrt{1 - \frac{{g}^2}{\pi}}}\,\gamma^5\,\tilde \phi^\prime (x) \Big \} \dt\,\psi_o (x)\,.
\ee

\no  The combination between the fields $\tilde\varphi$ and $\tilde\phi^\prime$ appearing in  Eqs. (\ref{e1}) and  (\ref{e2}) corresponds to a sine-Gordon field, whereas the combination appearing in Eq.(\ref{e3}) corresponds to a free massless field. This  suggest to perform  the following canonical field transformation,

\be\label{ct1}
\delta \tilde \Phi (x) = 
\sqrt \pi\,
\tilde \varphi (x)\,+\,\frac{g}{\sqrt{1\,-\,\frac{g^2}{\pi}}}\,\tilde \phi^\prime (x)\,,
\ee

\be\label{ct2}
\delta \tilde \xi (x) = \frac{g}{\sqrt{1\,-\,\frac{{g}^2}{\pi}}}\,\tilde \varphi (x) -
\sqrt \pi \,\tilde \phi^\prime (x)\,.
\ee

\no The value of the parameter $\delta$ is fixed by imposing canonical commutation relations for the fields $\tilde \Phi$ and $\tilde \xi$, 

\be
\frac{\delta^2}{\pi} = \frac{\beta^2}{4\,\pi} = \frac{1}{1 - \frac{{g}^2}{\pi}}\,.
\ee

\no The  fields $\tilde\phi^\prime$ and $\tilde\varphi$ can be written in terms of the new fields $(\tilde\Phi , \tilde\xi)$, as

\be
\tilde\phi^\prime (x) = \frac{g}{\sqrt\pi}\,\tilde\Phi (x)\,-\,\frac{\sqrt \pi}{\delta}\,\tilde\xi (x)\,,
\ee

\be
\tilde\varphi (x) = \frac{\sqrt\pi}{\delta}\,\tilde\Phi (x)\,+\,\frac{g}{\sqrt \pi}\,\tilde\xi (x)\,.
\ee

\no The equation of motion (\ref{e3}) is now

\be\label{ff}
\Box \tilde\xi (x) = 0\,,
\ee

\no and the vector current (\ref{e2}) of the MRS model is mapped into the current of the Thirring model,

\be\label{vcth}
{\cal J}_\mu (x)\,\equiv\,{\cal J}_\mu^{Th} (x)\,=\,-\, \frac{\beta}{2\pi}\,\epsilon_{\mu \nu}\,\partial^\nu\,\tilde \Phi (x)\,.
\ee

\no The mass operator (\ref{e1}) is identified with the mass operator of the Thirring model

\be
\vdt \bar \psi (x) \psi (x) \vdt \equiv \vdt \bar\Psi (x) \Psi (x) \vdt\,= - \frac{\mu}{\pi}\,\dt \cos\,\beta \,\tilde\Phi (x) \dt\,.
\ee

\no The hidden Thirring interaction in the MRS model is exhibited compactly in our operator approach. Using the fact that $\tilde\xi$ is a free and massless field,

\be
\varepsilon_{\mu \nu} \partial^\nu \tilde \xi (x) = \partial_\mu \xi (x)\,,
\ee

\no the Fermi field operator (\ref{psi}) can be rewritten in terms of the Wick-ordered exponential of the scalar field $\xi$,

\be\label{FB}
\psi (x)\,=\,{\cal Z}_\psi^{- \frac{1}{2}}\,\dt e^{\,i\,g \,\xi (x)}\dt\,\Psi (x)\,,
\ee

\no where  $\Psi$ is the Fermi field operator of the massive Thirring model given by the Mandelstam soliton operator,

\be
\Psi (x) = \Big (\frac{\mu}{2 \pi} \Big )^{1/2}\,
e^{\textstyle\,-\,i\,\frac{\pi}{4}\,\gamma^5}\,
\dt e^{\,\textstyle i\,\{\,\gamma^5\,\frac{\beta}{2}\,\tilde \Phi (x)\,+\,
\frac{2 \pi}{\beta}\,\int_{x^1}^{\infty}\,\partial_0\,{\tilde\Phi} (x^0,z^1)\,d z^1\,\}} \dt\,.
\ee

\no The field operator (\ref{FB}) corresponds to the Fermi field of the Thirring model interacting with the scalar field $\xi$ via vector-current-scalar-derivative coupling, i. e., the Schroer-Thirring model. The equation of motion (\ref{emf}) for the Fermi field  can be rewritten as,

\be\label{emff}
( i \gamma^\mu\,\partial_\mu\,-\,m_o\,)\,\psi (x)\,=\,
\,{g}^2\,\,N [\gamma^\mu\,\psi (x)\,{\cal J}_\mu^{Th}(x) ] \,+\,g\,
\,N  [\gamma^\mu\,\psi (x)\,\partial_\mu\,\xi (x) ] \,.
\ee

\no From the equations of motion  (\ref{emf}) and (\ref{emff}) we read off the equivalence between the MRS model with a massive fermion and the massive Thirring model with a vector-current-scalar-derivative 
interaction (Schroer-Thirring model) \footnote{The Schroer-Thirring model is defined by the classical Lagrangian density,

$$ 
{\cal L} (x) =
\bar \psi (x)\big ( i \gamma^\mu\,\partial_\mu\,-\,m_o\,\big )\,\psi (x) \,+\,\frac{1}{2}\,\partial_\mu\,\xi (x)\,\partial^\mu\,\xi (x)
$$

$$
+\,\frac{G^2}{2}\,
\Big (\bar\psi (x)\gamma^\mu\,\psi (x)\Big )\Big (\bar\psi (x)\gamma_\mu \psi (x) \Big ) \,+\,g\,
\,\Big (\bar\psi (x)\gamma^\mu\psi (x)\Big )\,\partial_\mu\,\xi (x)\,.
$$ 

\no The quantum theory is defined by the equations of motion

$$
\big ( i \gamma^\mu\,\partial_\mu\,-\,m_o\,\big )\,\psi (x)\,=\,G^2\,\vdt\,\gamma^\mu\,\psi (x)\,\Big (\,\bar\psi (x)\,\gamma_\mu\,\psi (x) \Big )\vdt\,+\,g\,\gamma^\mu\, N [\,\psi (x)\, \partial_\mu\, \xi (x)\,]\,,
$$

$$
\Box \xi (x)\,=\,-\,g\,\partial_\mu\,
\vdt \Big (\,\bar \psi (x)\,\gamma^\mu\,\psi (x)\,\Big )\vdt\,.
$$

\no The operator solution for the quantum equations of motion is given by Eq. (\ref{FB}) with 

$$
\beta^2 = \frac{4 \pi}{1 - \frac{G^2}{\pi}}\,.
$$

\no The Schroer model \cite{Schr} is obtained with $\beta^2 = 4\pi$ ($G = 0$). }. The equation of motion (\ref{emff}) corresponds to a particular case of the Schroer-Thirring model in which the Thirring coupling parameter and the derivative coupling parameter are the same. In this case the Thirring interaction cannot be turned off in order to give the Schroer model \cite{Schr}. The equivalence  (\ref{vcth}) implies that the Hilbert subspace of the MRS model generated by the correlation functions of the vector current ${\cal J}_\mu$ is isomorphic to the Hilbert subspace of the Thirring model generated by the correlation functions of the vector current ${\cal J}_\mu^{Th}$. The Wightman functions of the field operator $\psi (x)$ are those of the Fermi field $\Psi (x)$ of the Thirring model clouded by the contributions of the free massless field $\xi$,

$$
\langle 0 \vert \,\psi (x_1) \cdots \psi (x_n)\,\bar\psi (y_1) \cdots \bar\psi (y_n)\,\vert 0  \rangle \,=
$$

\be\label{Wf}
\langle 0 \vert \,\prod_{j = 1}^n\,\dt e^{\,\textstyle\,i\,g\,\xi (x_j)} \dt \,\prod_{k = 1}^n\,\dt e^{\textstyle\,-\,i\,
g\,\xi (y_k)} \dt \,\vert 0 \rangle\,\langle 0 \vert\,\Psi (x_1) \cdots \Psi (x_n)\,\bar\Psi (y_1) \cdots \bar\Psi (y_n)\,\vert 0  \rangle \,.
\ee

In view of Eq. (\ref{vcth}), the charge ${\cal Q}$ and the pseudo-charge ${\cal Q}^5$ carried by the Fermi field $\psi$ of the MRS model are mapped into the charges ${\cal Q}_{Th}$ and ${\cal Q}^5_{Th}$ carried by the Thirring field $\Psi (x)$

\be
[  {\cal Q}\,,\,\psi (x) ] = - \,\psi (x)\,\equiv\, [  {\cal Q}\,,\,\Psi (x) ] = - \Psi (x)\,,
\ee

\be
[  {\cal Q}^5\,,\,\psi (x) ] = \, - \, \gamma^5\,\Big ( \frac{1}{1 - \frac{g^2}{\pi}} \Big )\,\psi (x)\,\equiv\,[  {\cal Q}^5\,,\,\Psi (x) ]\,=\,-\, \gamma^5\,\frac{\beta^2}{4 \pi}\,\Psi (x)\,.
\ee

\no The charge sectors of the MRS model are mapped into the charge sectors of the massive Thirring model. The Hilbert space ${\cal H}$ of the MRS model is a direct product

\be
{\cal H} = {\cal H}_{\tilde\phi} \otimes {\cal H}_{\psi^{(0)}}\,,
\ee

\no with the selection rules

\be
{\cal Q} {\cal H}_{\tilde\phi} = 0\,\,\,\,,\,\,\,\,{\cal Q}^5 {\cal H}_{\tilde\phi} \neq 0\,,
\ee

\be
{\cal Q} {\cal H}_{\psi^{(0)}} \neq 0\,\,\,\,,\,\,\,\,{\cal Q}^5 {\cal H}_{\psi^{(0)}} \neq 0\,,
\ee

\no and is isomorphic to the Hilbert space

\be
{\cal H} = {\cal H}_\xi \otimes {\cal H}_\Psi
\ee

\no with the selection rules,

\be
{\cal Q} {\cal H}_{\xi} = 0\,\,\,\,,\,\,\,\,{\cal Q}^5 {\cal H}_{\xi} = 0\,,
\ee

\be
{\cal Q} {\cal H}_{\Psi} \neq 0\,\,\,\,,\,\,\,\,{\cal Q}^5 {\cal H}_{\Psi} \neq 0\,.
\ee

\no\hrulefill

It should be remarked that the Thirring current (\ref{vcth}) corresponds to the vector current of the Schroer-Thirring model defined with a regularization prescription for which the contribution of the scalar field $\xi$ is gauged away. Using the transformation (\ref{ct1})-(\ref{ct2}) into the current definition (\ref{cur}) one obtain,

$$
\vdt \, \bar\psi (x) \gamma^\mu \psi (x)\, \vdt _{RS}\,=\,
\lim_{\varepsilon \rightarrow 0}\,\Big \{\bar\psi (x + \varepsilon )\,\gamma^\mu\,e^{\,-\,i\,g^2\,\frac{\beta}{2 \pi}\,\int_x^{x + \varepsilon }\,\epsilon_{\mu \nu}\,\partial^\nu\,\tilde\Phi (z)\,d z^\mu
\,+\,i\,g\,\int_x^{x + \varepsilon }\,\partial_\mu\,\xi (z)\,d z^\mu
}\,\psi (x)\,-\,V.E.V.\Big \}
$$

\be\label{thpar}
=\,\lim_{\varepsilon \rightarrow 0}\,\Big \{\bar\Psi (x + \varepsilon )\,\gamma^\mu\,e^{\,-\,i\,g^2\,\frac{\beta}{2 \pi}\,\int_x^{x + \varepsilon }\,\epsilon_{\mu \nu}\,\partial^\nu\,\tilde\Phi (z)\,d z^\mu\,}\,\Psi (x)\,-\,V.E.V.\Big \}\,\equiv \vdt \,\bar \Psi (x) \gamma^\mu \Psi (x)\,\vdt _{Th}\,.
\ee

\no From (\ref{thpar}) we read off the appropriate regularization prescription for the computation of the vector current of the Thirring model. A general prescription for the current definition of the Schroer-Thirring model is given by

\be
\vdt\, \bar\psi (x) \gamma^\mu \psi (x)\, \vdt _{ST} =
\lim_{\varepsilon \rightarrow 0}\,\Big \{\bar\psi (x + \varepsilon )\,\gamma^\mu e^{\,-\,i\,G^2\,\frac{\beta}{2 \pi}\,\int_x^{x + \varepsilon }\,\epsilon_{\mu \nu}\,\partial^\nu\,\tilde\Phi (z)\,d z^\mu
\,-\,i\,a\,g\,\int_x^{x + \varepsilon }\,\gamma^5\,\epsilon_{\mu \nu}\,\partial^\nu\,\xi (z)\,d z^\mu
} \psi (x) - V.E.V. \Big \},
\ee
 
\no where $\underline{a}$ is an arbitrary parameter and $\psi (x)$ is given by (\ref{FB}) with

\be
\beta^2\,=\,\frac{4 \pi}{\sqrt{1\,-\,\frac{G^2}{\pi}}}\,.
\ee

\no Taking into account the fact that the model is scale invariant at small distances, the locality of the theory ensures the path independence of the line integral in the Mandelstam 
formula, which can be written as a line integral over a conserved current

\be
j^\mu (x) = \partial_\nu \,f^{\nu \mu} (x)\,,
\ee

\no with 

\be
f^{\nu \mu} (x)\, =\, -\, \epsilon^{\nu \mu}\,\tilde\Phi (x)\,,
\ee

\no and we obtain

\be
{\cal J}^\mu (x)_{ST} \,=\,-\,\frac{\beta}{2 \pi}\,\epsilon^{\mu \nu}\,\partial_\nu\,\tilde\Phi (x)\,-\,\frac{g}{2 \pi}\,(a + 1)\,\partial^\mu\,\xi (x)\,.
\ee

\no The Thirring current is obtained with $a = - 1$. The current corresponding to the Schroer model is recovered with $a = 1$ and $\beta^2 = 4 \pi$.

\section{Bosonized quantum Hamiltonian}
\setcounter{equation}{0}

In this section we obtain the complete bosonized Lagrangian of the MRS model. As in the case of the free massless fermion theory \cite{AAR} we shall first consider the bosonized quantum Hamiltonian. The bosonized composite operators of the quantum Hamiltonian are obtained as the leading operators in the Wilson short distance expansion for the operator products at the same point \cite{W}.

From the Lagrangian (\ref{L1}), the classical canonical momentum $\pi_{_{\tilde\phi}}$ conjugate to the field $\tilde\phi$ is formally given by the expression

\be
\pi_{_{\tilde\phi}} (x) = \partial^0\,{\tilde\phi} (x) + 
g\, \bar\psi (x)\,\gamma^0\,\gamma^5\,\psi (x)\,.
\ee

\no For $m_o = 0$, the quantum Hamiltonian density of the scale invariant model is obtained from the classical Hamiltonian with the classical fields replaced by their quantum operator counterparts and is given in terms of the normal-ordered operator products,

\be\label{ham}
{\cal H} (x) = \frac{1}{2}\,\dt \Big ( \partial_0 \tilde\phi (x) \Big )^2 \dt \,+\,\frac{1}{2}\,\dt \Big ( \partial_1 \tilde\phi (x) \Big )^2 \dt\,-\,
i\,\vdt\,\bar\psi (x)\,\gamma^1\,\partial_1\,\psi (x)\,\vdt\,-\,g\,\dt\,\Big ( \bar\psi (x)\,\gamma^1\,\gamma^5\,\psi (x) \Big )\,\partial_1 \tilde\phi (x)\,\dt\,,
\ee

\no with the axial current given by Eq. (\ref{5c}). In terms of the spinor components $\psi_\alpha$ ($\alpha = 1,2.$) the kinetic term of the Fermi field in the Hamiltonian (\ref{ham}) can be written as

\be
h (x) =\, - \,i\,\vdt\,\bar\psi (x)\,\gamma^1\,\partial_1\,\psi (x)\,\vdt\,=\,\sum_{\alpha = 1}^2\,( - 1 )^{\alpha + 1}\,h_\alpha (x)\,,
\ee

\no where,

\be\label{halpha}
h_\alpha (x)\,=\,i\,\vdt\,\psi^\dagger_\alpha (x)\,\partial_1\,\psi_\alpha (x)\,\vdt\,.
\ee

\no  We shall compute the composite operator $h_\alpha (x)$ as the leading term in the Wilson short distance expansion for the operator product at same point using the same regularization as that employed in the computation of the fermionic current. To begin with let us consider the point-splitting limit

\be\label{ih}
h_\alpha (x)\,=\,\lim_{\varepsilon \rightarrow 0}\,\Big \{\,h_\alpha (x ; \varepsilon )\,+\,h. c.\,-\,V. E. V. \Big \}\,,
\ee

\no where $h_\alpha ( x ; \varepsilon )$ is defined by the splitted operator product

\be\label{ha}
h_\alpha (x ; \varepsilon )\,=\,\frac{i}{2}\,\Bigg (\,\dt \psi^\dagger_\alpha (x + \varepsilon )\,
e^{\,- i\,g\,\int_{- \infty}^{x + \varepsilon}\,\epsilon_{\mu \nu}\,\partial^\nu\,\tilde\phi (z)\,d z^\mu} \dt \,\Bigg )\,\Bigg (\,\dt \,e^{\,i\,g\,\int_{- \infty}^{x}\,\epsilon_{\mu \nu}\,\partial^\nu\,\tilde\phi (z)\,d z^\mu}\,\partial_1\,\psi_\alpha (x) \dt\,\Bigg )\,.
\ee

\no With the operator solution (\ref{psi}), the operator product (\ref{ha}) can be written in terms of the Wick-ordered exponentials of the field $\tilde\phi$,

$$
h_\alpha (x ; \varepsilon )\,=\,{\cal Z}^{-1}_\psi (\varepsilon )\,\Bigg \{\,\Bigg (\,\dt e^{\,-\,i\,g\,\big ( \gamma^5_{\alpha\alpha}\,\tilde\phi (x + \varepsilon )\,+\,\int_{- \infty}^{x + \varepsilon}\,\epsilon_{\mu \nu}\,\partial^\nu\,\tilde\phi (z)\,d z^\mu\, \big )} \dt \,
\dt e^{\,i\,g\,\big ( \gamma^5_{\alpha\alpha}\,\tilde\phi (x)\,+\,\int_{- \infty}^{x}\,\epsilon_{\mu \nu}\,\partial^\nu\,\tilde\phi (z)\,d z^\mu \,\big )} \dt \Bigg )\,h_\alpha^{(0)} ( x ; \varepsilon )
$$

\be\label{ha1}
-\,\frac{g}{2}\,\gamma^5_{\alpha\alpha} \Bigg (
\dt e^{\,- i g \big ( \gamma^5_{\alpha\alpha} \tilde\phi (x + \varepsilon ) + \int_{- \infty}^{x + \varepsilon} \epsilon_{\mu \nu} \partial^\nu \tilde\phi (z) d z^\mu\, \big )} \dt
\dt e^{\,i g \big ( \gamma^5_{\alpha\alpha} \tilde\phi (x) + \int_{- \infty}^{x} \epsilon_{\mu \nu} \partial^\nu \tilde\phi (z) d z^\mu  \big )} \partial_1 \tilde\phi (x) \dt\,\Bigg )\psi_\alpha^{(0)^\dagger} (x + \varepsilon )\psi_\alpha^{(0)}(x) \Bigg \} ,
\ee

\no where $h_\alpha^{(0)} ( x ; \varepsilon )$ is the contribution of the kinetic term of the free Fermi field,

\be\label{h0}
h_\alpha^{(0)} ( x ; \varepsilon ) = \frac{i}{2}\,\psi_\alpha^{(0)^\dagger} (x + \varepsilon )\,\partial_1\,\psi_\alpha^{(0)} (x)\,.
\ee

\no In the computation of (\ref{ha1}) we shall use that,

\be
\Big (\,\gamma^5_{\alpha\alpha}\,\varepsilon^\mu\,\partial_\mu\,+\,\varepsilon^\mu\,\epsilon_{\mu \nu} \partial^\nu \Big )\,\tilde\phi (x) = \mp\,\varepsilon^\pm\,\partial_\pm \tilde\phi (x)\,\,\,,\,\,\alpha = 1 , 2\,,
\ee

\no and that if $ [ B , A ] = $ c - number,

\be\label{AB}
e^{-\,B}\,A\,=\,A\,e^{-\,B}\,-\,[ B , A ]\,e^{-\,B}\,,
\ee

$$
\Big (\,\dt e^{\,-\,i\,a\,\Phi (x)} \dt \,\Big )\,\Big (\, \dt e^{\,i\,a\,\Phi (y)}\,\partial_1\,\Phi (y) \dt \Big )\,=
$$

\be
e^{\,a^2\,D^{(+)} (x - y)}\,\Bigg \{\,\dt e^{\,-\,i\,a\,[\Phi (x)\,-\,\Phi (y)]}\,\partial_1\,\Phi (y) \dt \,-\,
i\,a\,\Big (\,\partial_{y^1}\,D^{(+)} (x - y)\,\Big )\,
\dt e^{\,-\,i\,a\,[\Phi (x)\,-\,\Phi (y)]} \dt\,\Bigg \}\,,
\ee

\no where

\be
D^{(+)} (x) = [ \Phi^{(+)} (x)\,,\,\Phi^{(-)} (0) ]\,.
\ee

\no Performing the normal ordering of the exponentials of the field $\tilde\phi$ we can decompose (\ref{ha1}) as follows,

\be\label{h}
h_\alpha (x ; \varepsilon )\,=\,h^{(I)}_\alpha (x ; \varepsilon )\,+\,h^{(II)}_\alpha (x ; \varepsilon )\,+\,h^{(III)}_\alpha (x ; \varepsilon )\,.
\ee

\no where

\be\label{hI}
h^{(I)}_\alpha (x ; \varepsilon )\,=\,\dt e^{\,\pm\,i\,g\,\varepsilon^\pm\,\partial_\pm \tilde\phi (x)} \dt \,h_\alpha^{(0)} (x ; \varepsilon )\,,
\ee

\be\label{hII}
h^{(II)}_\alpha (x ; \varepsilon )\,=\,-\,\frac{g}{2}\,\Big ( \psi_\alpha^{(0)^\dagger} (x + \varepsilon )\psi_\alpha^{(0)}(x) \Big )\,\dt e^{\,\pm\,i\,g\,\varepsilon^\pm\,\partial_\pm \tilde\phi (x)}\,\partial_1 \tilde\phi (x) \dt\,,
\ee

\be\label{hIII}
h^{(III)}_\alpha (x ; \varepsilon )\,=\,i\,\frac{g^2}{2}\,\gamma^5_{\alpha\alpha}\,F_\alpha (\varepsilon )\,\Big (
\psi_\alpha^{(0)^\dagger} (x + \varepsilon )\psi_\alpha^{(0)}(x) \Big )\,\dt e^{\,\pm\,i\,g\,\varepsilon^\pm\,\partial_\pm \tilde\phi (x)} \dt \,,
\ee

\no where the singular function $F_\alpha ( \varepsilon )$ is given by the commutator

\be\label{f}
F_\alpha (\varepsilon ) = \Big [ \gamma^5_{\alpha\alpha}\,\tilde\phi^{(+)} (x  + \varepsilon ) + \int_{- \infty}^{x + \varepsilon}\,\epsilon_{\mu \nu} \partial^\nu\,\tilde\phi^{(+)} (z) d z ^\mu\,,\,\partial_1\,\tilde\phi^{(-)} (x) \Big ]\,=\,\frac{-1}{2 \pi \varepsilon^\pm}\,.
\ee

Let us consider the term $h^{(I)}$. In order to compute the free field contribution $h^{(0)} (x)$, given by Eq. (\ref{h0}), we shall make use of the following relations

\be\label{u1}
\varepsilon^\mu \partial_\mu\,\varphi_{\ell,r} (x) =\,\mp\,\frac{1}{2}\,\varepsilon^\pm\,\partial_\pm  \tilde\varphi (x)\,,
\ee

\be
\partial_1 \varphi_{\ell,r} (x) = -\,\frac{1}{2}\,\partial_\pm \tilde\varphi (x)\,,
\ee

\no corresponding to $\alpha = 1 , 2$, respectively. Using (\ref{AB}) and normal ordering the exponentials of the field $\tilde\varphi$, the free fermion contribution is given by \cite{AAR},

$$
h_\alpha^{(0)} (x ; \varepsilon ) =\,-\,\frac{i}{4 \sqrt\pi\,\varepsilon^\pm}\,\dt e^{\,\pm\,i\,\sqrt\pi\,\varepsilon^\pm\,\partial_\pm\,\tilde\varphi (x)}\,\partial_\pm \tilde\varphi (x)\dt
$$

\be\label{18}
+\,\big (\frac{1}{\varepsilon^\pm}\big )\,[ \varphi_{\ell,r}^{(+)} (x + \varepsilon )\,,\,\partial_1\,\varphi_{\ell,r}^{(-)} (x) ]\,\dt e^{\,\pm\,i\,\sqrt\pi\,\varepsilon^\pm\,\partial_\pm\,\tilde\varphi (x)} \dt\,,
\ee

\no where

\be\label{16}
[ \varphi_{\ell,r}^{(+)} (x + \varepsilon )\,,\,\partial_1\,\varphi_{\ell,r}^{(-)} (x) ]\,=\,\frac{\pm\,1}{4 \pi\,\varepsilon^\pm}\,.
\ee

\no  Expanding the first exponential in (\ref{18}) in powers of $\varepsilon$ up to first order and the second exponential up to second order, we obtain

\be\label{h00}
h_\alpha^{(0)} ( x ; \varepsilon ) =\,\pm\,\frac{1}{8}\,\dt \Big (\,\partial_\pm\,\tilde\varphi (x) \Big )^2 \dt \,\pm\,\frac{1}{4 \pi\,(\varepsilon^\pm )^2}\,+\,{\cal O} (\varepsilon )\,.
\ee

\no Introducing (\ref{h00}) into (\ref{hI}), in order to compute the leading operator in 
the $\varepsilon$-expansion of $h^{(I)} ( x ; \varepsilon )$  we need retain terms up to 
second order in $\varepsilon$ in the exponentials, 

$$
h_\alpha^{(I)} (x) = h^{(I)}_\alpha (x ; \varepsilon ) + h.c. - V.E.V. =
$$

\be
\pm\,\frac{1}{8}\,\dt \Big (\,\partial_\pm\,\tilde\varphi (x) \Big )^2 \dt \,\mp\,\frac{1}{8}\,\Big ( \frac{g^2}{\pi} \Big )\,\dt \Big (\,\partial_\pm\,\tilde\phi (x) \Big )^2 \dt\,.
\ee

\no Combining the contributions from the two spinor components, the leading operator $h^{(I)} (x)$ is given by

$$
h^{(I)} (x) = h_1^{(I)}(x) - h_2^{(I)} (x)
$$

\be\label{fhI}
 = \frac{1}{2}\,\Big \{\,\dt \Big (\partial_0 \tilde\varphi (x) \Big )^2 \dt \, +\, \dt \Big (\partial_1 \tilde\varphi (x) \Big )^2 \dt \,\Big \}\, -\, \frac{1}{2}\,\Big (\frac{g^2}{\pi}\Big )\,\Big \{\,\dt \Big (\partial_0 \tilde\phi (x) \Big )^2 \dt \,+\, \dt \Big (\partial_1 \tilde\phi (x) \Big )^2 \dt\,\Big \}\,.
\ee

\no The first term in (\ref{fhI}) corresponds to the bosonized Hamiltonian of the free massless Fermi field \cite{AAR}. The term proportional to $g^2$ in (\ref{fhI}) is the quantum correction to the free Lagrangian piece of the field $\tilde\phi$.

Now let us consider the second term $h^{(II)}$. To this end we must compute the operator product of the free fermion field appearing in Eq. (\ref{hII}). Using (\ref{u1}) and normal ordering the exponential, one obtain

\be\label{u2}
\psi^{(0)^\dagger}_\alpha (x + \varepsilon ) \psi^{(0)}_\alpha (x)\,=\,\frac{1}{2 i \pi\,\varepsilon^\pm}\,\dt e^{\,\pm\,i\,\sqrt\pi\,\varepsilon^\pm\,\partial_\pm\,\tilde\varphi (x)} \dt\,.
\ee

\no Introducing the pseudoscalar potential $\widetilde{\cal J}$,

\be\label{J}
\widetilde{\cal J} (x) = \frac{1}{\sqrt\pi}\,\tilde\varphi (x) + \frac{g}{\pi}\,\tilde\phi (x)\,,
\ee

\no such that the axial-current can be written as

\be
{\cal J}^5_\mu (x) =\,-\,\partial_\mu \widetilde{\cal J} (x)\,,
\ee

\no and using (\ref{u2}),  we can writte (\ref{hII}) as

\be\label{II}
h_\alpha^{(II)} (x ; \varepsilon ) = \frac{g}{4 \pi}\,\gamma^5_{\alpha\alpha}\,\Big (\frac{i}{\varepsilon^\pm} \Big )\,
\dt e^{\,\pm\,i\,\pi\,\varepsilon^\pm\,\partial_\pm\,\widetilde{\cal J} (x)}\,\partial_1 \tilde\phi (x) \dt\,.
\ee

\no Expanding the exponential in (\ref{II}) in powers of $\varepsilon$ up to first order, one has

\be
h_\alpha^{(II)} (x) = h^{(II)}_\alpha (x ; \varepsilon ) + h.c. - V.E.V. =\,\pm\,\frac{g}{\pi}\,\dt \Big (\,\partial_\pm \widetilde{\cal J} (x)\,\partial_1 \tilde\phi (x) \Big ) \dt \,+\,{\cal O} (\varepsilon )\,.
\ee

\no The leading operator in the second contribution $h^{(II)} (x) = h^{(II)}_1 (x) - h^{(II)}_2 (x)$ is then given by,

\be\label{fhII}
h^{(II)} (x) =\,-\,g\,\dt \partial^1 \widetilde{\cal J} (x)\,\partial_1 \tilde\phi (x) \dt\,=\,g\,\dt \Big (\,\bar \psi (x) \gamma^1 \gamma^5 \psi (x)\Big )\,\partial_1 \tilde\phi (x) \dt \,.
\ee

\no As expected from the operator solution (\ref{psi}), the contribution (\ref{fhII}) cancels the corresponding term in the Hamiltonian (\ref{ham}).

Finally, let us consider the term $h^{(III)}$. Using (\ref{u2}), (\ref{J}) and (\ref{f}), we can write (\ref{hIII}) as follows 

\be
h_\alpha^{(III)} (x ; \varepsilon )\,=\,\pm\,\Big ( \frac{g^2}{8\pi^2} \Big )\,\frac{1}{(\varepsilon^\pm )^2}\,\dt e^{\,\pm\,i\,\pi\,\varepsilon^\pm\,\partial_\pm\,\widetilde{\cal J} (x)} \dt\,.
\ee

\no Expanding the exponential in powers of $\varepsilon$ up to second order, we get

\be
h_\alpha^{(III)} (x ; \varepsilon ) \,=\,\mp\,\frac{g^2}{16}\,\dt \Big ( \partial_\pm \widetilde{\cal J} (x) \Big )^2 \dt\,+\,g^2\,\Big (\,\frac{i}{8\,\pi\,\varepsilon^\pm}\,\Big )\,\partial_\pm \widetilde{\cal J} (x)\,\pm\,\frac{g^2}{8 \pi^2 (\varepsilon^\pm)^2}\,+\,{\cal O} (\varepsilon )\,.
\ee

\no One obtains

\be
h^{(III)}_\alpha (x) = h_\alpha^{(III)} (x ; \varepsilon ) + h.c. - V.E.V. = \mp\,\frac{g^2}{8}\,\dt \Big ( \partial_\pm \widetilde{\cal J} (x) \Big )^2 \dt\,.
\ee

\no The leading term $h^{(III)} (x)$ is given by

\be\label{fhIII}
h^{(III)} (x) \,=\,-\,\frac{g^2}{4}\,\Big \{\,\dt \Big (\partial_0 \widetilde{\cal J} (x) \Big )^2 \dt\,+\,
\dt \Big (\partial_1 \widetilde{\cal J} (x) \Big )^2 \dt\,\Big \}\,,
\ee

\no and corresponds to the contribution of the Thirring interaction to the 
quantum Hamiltonian. Collecting all terms (\ref{fhI})-(\ref{fhII})-(\ref{fhIII}), the bosonized form of the fermionic kinetic term (\ref{ih}) is given by

$$
h (x)\,=\,i\,\vdt\,\bar\psi (x) \gamma^1 \partial_1 \psi (x)\,\vdt\,=
\frac{1}{2} \dt\Big (\partial_0 \tilde\varphi (x) \Big )^2 \dt\, +\, \frac{1}{2}\dt \Big (\partial_1 \tilde\varphi (x) \Big )^2 \dt\, 
$$

$$
-\, \frac{1}{2}\,\Big (\frac{g^2}{\pi}\Big )\,\Big \{\,\dt \Big (\partial_0 \tilde\phi (x) \Big )^2 \dt\, +\, \dt \Big (\partial_1 \tilde\phi (x) \Big )^2 \dt\,\Big \}
$$

\be
+\,g\,\dt \Big (\,\bar \psi (x) \gamma^1 \gamma^5 \psi (x)\Big )\,\partial_1 \tilde\phi (x) \dt
\,-\,\frac{g^2}{4}\,\Big \{\,\dt \Big (\partial_0 \widetilde{\cal J} (x) \Big )^2 \dt\,+\,
\dt \Big (\partial_1 \widetilde{\cal J} (x) \Big )^2 \dt \,\Big \}\,.
\ee

\no Introducing the mass perturbation the total bosonized quantum Hamiltonian (\ref{ham}) is given by

$$
{\cal H}_{bos} (x) \,=\,
\frac{1}{2} \dt\big(\partial_0 \tilde\varphi (x) \big)^2 \dt\, +\, \frac{1}{2} \dt \big(\partial_1 \tilde\varphi (x) \big)^2 \dt\, \,+\,\frac{1}{2}\,\Big ( 1\,-\,\frac{g^2}{\pi}\Big )\,\Big \{\,\dt \big(\partial_0 \tilde\phi (x) \big)^2 \dt \, +\, \dt \big(\partial_1 \tilde\phi (x) \big)^2 \dt\,\Big \}
$$

\be
-\,\frac{g^2}{4}\,\Big \{\,\dt \big (\partial_0 \widetilde{\cal J} (x) \big )^2 \dt\,+\,
\dt \big (\partial_1 \widetilde{\cal J} (x) \big )^2 \dt\,\Big \}\,+\,m^\prime_o\,\dt \cos \{2 \sqrt \pi\,\tilde\varphi (x) + 2 g \tilde\phi (x) \} \dt\,.
\ee

\no where $m^\prime_o = \mu\,m_o / \pi$. The corresponding bosonized Lagrangian density is 

$$
{\cal L}_{bos} (x)\,=\,\frac{1}{2} \dt \partial^\mu \tilde\varphi (x)\,\partial_\mu \tilde\varphi (x) \dt\,
+\,\frac{1}{2}\,\Big ( 1\,-\,\frac{g^2}{\pi}\Big )\,\dt \partial^\mu\,\tilde\phi (x)\,\partial_\mu \tilde\phi (x) \dt\,
$$

\be\label{boslag}
-\,\frac{g^2}{4}\,\dt \partial_\mu\,\widetilde{\cal J}(x)\,\partial^\mu\,\widetilde{\cal J} (x) \dt \,-\,m^\prime_o\,\dt \cos \{2 \sqrt \pi\,\tilde\varphi (x) + 2 g \tilde\phi (x) \} \dt\,,
\ee

\no with $\tilde{\cal J}$ given by (\ref{J}). In the bosonized theory the new momentum $\Pi_{_{\tilde\phi}}$ conjugate to the field $\tilde\phi$ is given by

\be
\Pi_{_{\tilde\phi}} (x)\, =\,\Big ( 1\,-\,\frac{g^2}{\pi} \Big )\, \partial_0\,\tilde\phi (x)\,-\,\big ( \frac{g}{\pi}\big )\,\frac{g^2}{2}\,\partial_0\,\widetilde{\cal J} (x)\,,
\ee

\no and the momentum $\Pi_{_{\tilde\varphi}}$ associated with the field $\tilde\varphi$ is

\be
\Pi_{_{\tilde\varphi}} (x)\, =\, \partial_0\,\tilde\varphi (x)\,-\,\big (\frac{1}{\sqrt\pi}\big )\,\frac{g^2}{2}\,
\partial_0\,\widetilde{\cal J} (x)\,.
\ee

\no Thus, one has

\be
\dt \Pi_{_{\tilde\varphi}} (x)\,\partial_0\,\tilde\varphi (x) \dt\,+\,\dt \Pi_{_{\tilde\phi}} (x)\,\partial_0\,\tilde\phi (x) \dt\, =\,
\dt \Big (\partial_0 \tilde\varphi (x) \Big )^2 \dt\,+\,\Big ( 1 - \frac{g^2}{\pi} \Big )\,\dt \Big (\partial_0 \tilde\phi (x) \Big )^2 \dt\,-\,\frac{g^2}{2}\,\dt \Big ( \partial_0 \tilde{\cal J} (x) \Big )^2 \dt\,.
\ee

\no From the bosonized Lagrangian we obtain the following coupled equations of motion,

\be\label{eq1}
\Box\,\tilde\varphi (x)\,-\,\Big (\frac{g^2}{2}\Big )\,\frac{1}{\sqrt\pi}\,\Box\,\Big ( \frac{1}{\sqrt\pi}\,\tilde\varphi (x)\,+\,\frac{g}{\pi}\,\tilde\phi (x) \Big )\,=\,2\,\sqrt\pi\,m_o^\prime\,\dt \sin\,\Big ( 2 \sqrt\pi\,\tilde\varphi (x)\,+\,2 g\,\tilde\phi (x) \Big ) \dt\,,
\ee 

\be\label{eq2}
\Big ( 1 - \frac{g^2}{\pi} \Big )\,\Box\,\tilde\phi (x)\,-\,\Big (\frac{g^2}{2}\Big )\,\frac{g}{\pi}\,\Box\,\Big ( \frac{1}{\sqrt\pi}\,\tilde\varphi (x)\,+\,\frac{g}{\pi}\,\tilde\phi (x) \Big )\,=\,2\,g\,m_o^\prime\,\dt \sin\,\Big ( 2 \sqrt\pi\,\tilde\varphi (x)\,+\,2 g\,\tilde\phi (x) \Big ) \dt\,,
\ee 

\no in agreement with the bosonized equation of motion (\ref{bem}). Rescaling the field $\tilde\phi$ by (\ref{fs}) and performing the canonical transformation (\ref{ct1})-(\ref{ct2}), the bosonized Lagrangian density (\ref{boslag}) can be rewritten as,

$$
{\cal L}_{bos} (x)\,=\,\frac{1}{2} \dt \partial^\mu \tilde\xi (x)\,\partial_\mu \tilde\xi (x) \dt\,
+\,\frac{1}{2}\,\dt \partial^\mu\,\tilde\Phi (x)\,\partial_\mu \tilde\Phi (x) \dt\,
$$

\be\label{bosl}
+\,\frac{g^2}{4}\,\dt \Big ( \frac{\beta}{2\pi}\,\epsilon_{\mu\nu}\partial^\nu\,\tilde\Phi(x)\Big )\,\Big (\frac{\beta}{2\pi}\,\epsilon^{\mu\rho}\partial_\rho\,\tilde\Phi (x)\Big ) \dt \,-\,m^\prime_o\,\dt \cos \beta \,\tilde\Phi (x) \dt\,.
\ee
\\

\no The Lagrangian (\ref{bosl}) corresponds to the Lagrangian of the Thirring model and a ``decoupled'' free massless field. Although the field $\tilde\xi$ decouples in the bosonized Lagrangian and thus the partition function factorizes 

\be
{\LARGE{Z}} [0] =  {\Large{Z}}_\xi [0]\,\times \, {\Large{Z}}_{\tilde\Phi} [0]\,,
\ee

\no the field $\xi$ does not decouple in the generating functional and gives contribution to the Wightman functions of the Fermi field (\ref{Wf}).
\section{Concluding Remark}

Using the operator formulation, we have analyzed the Rothe-Stamatescu model with a massless pseudoscalar field interacting with a massive Fermi field (MRS model). The hidden Thirring interaction in the MRS model is exhibited compactly by performing a canonical transformation on the Bose field algebra. The present approach give us easiness to read off the equivalence between the MRS model and the Schroer-Thirring model, i. e., the Thirring model with an additional 
vector-current-scalar derivative interaction. The isomorphism is established for a particular case of the Schroer-Thirring model for which the Thirring coupling parameter and the derivative coupling parameter are the same ($G^2 = g^2$). The bosonized quantum Hamiltonian were obtained as the leading operator in the $\varepsilon$-expansion
for the operator products at the same point. 

One may conjecture that the intrinsic property of the MRS model of being equivalent to the Schroer-Thirring model should be identified as the underneath phenomena that enable one to establish the weak equivalence between the Thirring model and the fermionic sector of the derivative coupling model with two derivative coupling interactions, as proposed in Refs. \cite{4,5,6}. However, at the operator level several aspects of the weak equivalence between the two models, as obtained in Ref. \cite{5}, still remain obscure and merit a separate investigation within the present operator approach. 

As stressed in Ref. \cite{Schr}, fermions coupled to zero rest mass particles may not be eigenstates of the mass operator. The concept of `` infraparticles '' was introduced by Schroer in Ref. \cite{Schr}, which is a particle interacting with a massless scalar field that loses its discrete spectrum owing to infrared 
radiation. It should be very interesting to investigate in the present case whether the Thirring interaction prevents, or not, the existence of the infraparticle and the gap situation can be restored.

{\bf Acknowledgments}: The authors are greateful to Brazilian Research Council (CNPq) for partial financial support. The authors wish to thank K. D. Rothe for helpful comments and suggestions.






\begin{thebibliography}{12}
\bibitem{Schr} B. Schroer, Fortschritte der Physik {\bf 11} (1963) 1.

\bibitem{RS} K. D Rothe and I. O. Stamatescu, Ann. of Phys. (NY) {\bf 95} (1975) 202.

\bibitem{1} M. El Afioni, M. Gomes and R. K\"oberle, Phys. Rev. D {\bf 19} (1979) 1144.

\bibitem{2} M. El Afioni, M. Gomes and R. K\"oberle, Phys. Rev. D {19} (1979) 1791.

\bibitem{3} Carlos Farina and Arvind Vaidya, Phys. Rev. D {\bf 32} (1985) 2243.

\bibitem{4} A. J. da Silva, M. Gomes and R. K\"oberle, Phys. Rev. D {\bf 34} (1986) 504.

\bibitem{5} M. Gomes and A. J. Silva, Phys. Rev. D {\bf 34} (1986) 3916.

\bibitem{6} R. Banerjee, Phys. Rev. D {\bf 37} (1988) 3778.

\bibitem{7} T. Ikehashi, Phys. Lett. {\bf B 313} (1993) 103.

\bibitem{bel} L. V. Belvedere, J. Phys. {\bf A} {\bf 33} (2000) 2755.

\bibitem{LS} S. Lowenstein and J. A. Swieca, Ann. of Phys. {\bf 68} (1971) 172.

\bibitem{bel1} L. V. Belvedere, J. A. Swieca, K. D. Rothe and B. Schroer, Nucl. Phys.                 {\bf B 153} (1979) 112; L. V.Belvedere, Nucl. Phys. {\bf B 276} (1986)                 197.

\bibitem{Wightman} A. S. Wightman, in ``{\it Carg\`ese Lectures in Theoretical                           Physics}'', (1964), ed. M. Levy, Gordon and Breach, New York, 1966.

\bibitem{Swieca} J. A. Swieca, Fortschritte der Physik {\bf 25} (1977) 303.

\bibitem{AAR} E. Abdalla, M. C. Abdalla and K. D. Rothe, {\it Non-Perturbative Methods               in $2$-Dimensional Quantum Field Theory}, 1991, (Singapore: World Scientific ).

\bibitem{M} S. Mandelstam, Phys. Rev. {\bf D 11} (1975) 3026.

\bibitem{W} K. Wilson, Phys. Rev. {\bf 179} (1969) 1499. 
\end{thebibliography}
\end{document}